\UseRawInputEncoding

\documentclass[aps,prd,nofootinbib,amsmath,amssymb,superscriptaddress,showpacs,preprintnumbers,twocolumn,10pt]{revtex4}

\usepackage{graphics,epsfig,subfigure}
\usepackage{ulem}
\usepackage{color}
\usepackage{url}
\usepackage{float}
\usepackage{dcolumn}
\usepackage{bm}
\usepackage{amssymb}
\usepackage{latexsym}
\usepackage{booktabs}
\usepackage{amsmath}
\allowdisplaybreaks[4]
\usepackage{multirow}
\usepackage[colorlinks=true, linkcolor=red, citecolor=blue]{hyperref}

\newcommand{\be}{\begin{equation}}
\newcommand{\ee}{\end{equation}}
\newcommand{\bq}{\begin{eqnarray}}
\newcommand{\eq}{\end{eqnarray}}

\begin{document}
\newcommand{\PR}[1]{\ensuremath{\left[#1\right]}} 
\newcommand{\PC}[1]{\ensuremath{\left(#1\right)}} 
\newcommand{\PX}[1]{\ensuremath{\left\lbrace#1\right\rbrace}} 
\newcommand{\BR}[1]{\ensuremath{\left\langle#1\right\vert}} 
\newcommand{\KT}[1]{\ensuremath{\left\vert#1\right\rangle}} 
\newcommand{\MD}[1]{\ensuremath{\left\vert#1\right\vert}} 

\title{Gravitational lensing by a black hole in effective loop quantum gravity}
\author{Qi-Ming Fu}
\email{fuqiming@snut.edu.cn}
\affiliation{Department of Physics, College of Sciences, Northeastern University, Shenyang 110819, China}
\affiliation{Institute of Physics, Shaanxi University of Technology, Hanzhong 723000, China}

\author{Xin Zhang\footnote{Corresponding author}}
\email{zhangxin@mail.neu.edu.cn}
\affiliation{Department of Physics, College of Sciences, Northeastern University, Shenyang 110819, China}
\affiliation{Key Laboratory of Data Analytics and Optimization for Smart Industry (Northeastern University), Ministry of Education, China}



\begin{abstract}
It is well known that general relativity is an effective theory of gravity at low energy scale, and actually quantum effects cannot be ignored in the strong-field regime. As a strong gravitational object, black hole plays a key role in testing the quantum effects of gravity in the strong-field regime. In this paper, we focus on black hole in effective loop quantum gravity and investigate what the influences are of the quantum effects on the weak and strong bending angles of light rays. We find that this black hole could be a Schwarzschild black hole, a regular black hole, a one-way traversable wormhole, or a two-way traversable wormhole for the different values of the quantum parameter, and the strong bending angle for this compact object exhibits two different divergent behaviors, i.e., the logarithmic divergence and non-logarithmic divergence. There are a series of relativistic images on both sides of the optical axis. Only the outermost one can be resolved as a single image, and all others are packed together at the limiting angular position. It is interesting to note that the angular separation between the outermost relativistic image and the others initially increases and then decreases as the quantum parameter increases, indicating that there is a maximum in the angular separation. The maximum is reached after the black hole becomes a wormhole, which can be taken as a signal for the formation of the wormhole. Moreover, the limiting angular position decreases as the quantum parameter increases but a little bounce occurs after the formation of the wormhole, and the relative magnification in magnitudes first decreases and then increases as the quantum parameter increases.
\end{abstract}





\maketitle

\section{Introduction}

It is widely known that general relativity (GR) has been very successful in modern physics. It has passed an array of tests on  solar systems \cite{Will4} as well as on binary pulsars \cite{Will4,Wex102} and cosmology \cite{Clifton1,Ferreira335}. Besides, the direct detections of gravitational waves \cite{Abbott061102,Abbott141101,Abbott161101} and the image of the black hole shadow \cite{AkiyamaL1} further confirm the validity of GR at the scale of high curvature, such as near neutron stars or black holes. However, GR may not be the complete theory of gravity, since its nonrenormalization, the disability of explaining dark matter and dark energy, and so on. One of the most important issues is the prediction of the spacetime singularity inside a black hole or at the beginning of the universe, which suggests that the theory breaks down there. Usually, it is believed that the singularity can be circumvented by some quantum effects. To address the issues like this, one need to resort to the quantum theory of gravity.

Although a well-developed quantum theory of gravity is still out of reach, many efforts have been made to understand the spacetime singularity. One of the most successful attempts is the phase space quantization, i.e., the so-called polymerization schemes developed in loop quantum gravity (LQG), which has been used to resolve the big-bang singularity \cite{Bojowald4,Singh125005,Zhang661,Ashtekar213001}. In this quantization scheme, a small parameter called the polymer scale is introduced. When approaching this scale, the quantum effects can no longer be neglected. On the other hand, as an important candidate of the strong gravitational regime, black hole plays a key role in understanding the nature of spacetime at high-energy scales. It is natural to describe the black-hole spacetimes by considering the quantum corrections, such as the polymerization schemes. Over the past few years, a lot of effective polymerized black holes have been constructed \cite{Modesto5587,Campiglia3649,Caravelli245022,Barbero02963,Perez126901,Corichi055006,Olmedo225011,Ashtekar241301,Bodendorfer195015,Liu084001,Kelly106024,Sartini066014,Bodendorfer136390,Bodendorfer095002,Gambini205012}, most of which focus on the spherically symmetric cases. The common features of these black holes are that the classical singularities inside them are replaced by a transition surface, which connects a black hole to a white hole, and that the spacetime is regular everywhere. Besides, since the resulting quantum corrected spacetimes can still be described by an explicit line element, one can study such nonsingular effective geometries by using the well-developed mechanism for semiclassical black holes.

There have already been some investigations on various aspects of the polymerized black holes in the literature. For example, {the authors of Ref.~\cite{Alesci1656} studied the particle production by a regular LQG black hole, and showed that the unitarity is recovered for the black hole evaporation. The relation between the polymerized black holes and the non-singular black holes in mimetic gravity was investigated in Ref.~\cite{Achour072}. The exact solutions for the static spherically symmetric black hole interior were obtained in Ref.~\cite{Achour20006}. In a status report \cite{Barrau102}, the authors gave a review about the possible experimental or observational consequences of black holes in LQG. The greybody factors for the scalar and fermion fields in the background spacetime of a polymerized black hole were explicitly calculated in Ref.~\cite{Moulin125003}. The quasinormal modes of the polymerized spherically symmetric black holes were studied in Refs.~\cite{Moulin202,Bouhmadi066}.} In Ref.~\cite{Mele04788}, the authors analyzed the thermodynamic properties of the quantum corrected polymerized black holes and investigated the effects of the corresponding quantum corrections as functions of the masses of the black and white holes. {The Hawking radiation spectra and evaporation of spherically symmetric black holes in LQG were investigated in Refs.~\cite{Barrau251301,Gambini115003,Barrau124046,Ashtekar21,Arbey104010,Arbey084016}.} In Ref. \cite{Brahma181301}, the authors constructed the rotating polymerized black hole using the revised Newman-Janis algorithm and investigated the shadow cast by this black hole. One can see Refs.~\cite{Achour041,Achour124041,Achour020,Gan124030,Blanchette084038,Munch046019,Liu02861} for more investigations on the polymerized black holes. In this paper, we mainly focus on the polymerized black hole proposed in Refs.~\cite{Bodendorfer136390,Bodendorfer095002} and investigate the strong and weak deflection angles of light rays by this black hole.

On the other hand, light rays will be bent as they pass through a massive object, which is known as gravitational lensing, and is a prediction of GR. Since the first observation of the deflection of starlight near the Sun in 1919, gravitational lensing by other astronomical objects has been observed many times. In modern astronomy, gravitational lensing plays a key role in many aspects in astrophysical and cosmological observations. Most studies of gravitational lensing are in the weak deflection limit, and the bending angle is at most a few arcseconds. However, when passing close to a black hole, light rays may revolve many times around the black hole before arriving at the observer. This phenomenon is known as the so-called strong gravitational lensing, which was introduced a few decades ago for Schwarzschild spacetime \cite{Darwin180}. An analytical expression for the strong deflection angle of the Schwarzschild black hole was given in Ref.~\cite{Ohanian428}. Subsequently, the observables, such as the positions, magnifications, and time delays of the images for the Schwarzschild black hole in the strong deflection limit were investigated systematically in Refs.~\cite{Frittelli064021,Virbhadra084003,Bozza1535,Virbhadra083004}. In Ref.~\cite{Bozza103001}, the author introduced an analytical framework to investigate the strong gravitational lensing effect for a generic static spherically symmetric metric, and then applied it to Reissner-Nordstr$\text{\"o}$m black hole and other spherically symmetric objects with a photon sphere. Numerical investigations on the gravitational lensing were performed in Refs.~\cite{Virbhadra084003,Virbhadra083004,Virbhadra124014}. Other interesting studies on gravitational lensing in spherically symmetric spacetimes can be found in Refs.~\cite{Eiroa024010,Bhadra103009,Mukherjee583,Gyulchev023006,Liu124017,Eiroa085008,Bin-Nun123011,Eiroa083009,Tsukamoto104062,Tsupko124009,Wei253,Zhao007,Tsukamoto124001,Chakraborty045,Shaikh044037,Nascimento044021,Chagoya075026,Furtado044047,Takizawa104039,Gao136683,Eiroa043002,Chen024036,Tsukamoto024033,Tsukamoto064022,Parvizi03069}. Besides, gravitational lensing by a rotating black hole has been analyzed in Refs.~\cite{Bozza103006,Bozza083003,Kraniotis085021,Islam124052}.

Although there are many studies on polymerized black holes, the deflection of light rays by this type of black holes has not yet been studied. The deflection angle of light rays can be used as an effective way to study black holes and can provide us with valuable information about black holes. For example, the mass of the black hole and the quantum parameters resulting from quantum effects can be determined by analyzing the angular position of the relativistic images. In addition, the angular separation between the outermost relativistic image and the other images can present interesting features. Thus, it is necessary to study the gravitational lensing effect of a polymerized black hole in detail, and it is interesting to know what the effects are of the quantum corrections on the deflection angle.

This paper is organized as follows. In Sec.~\ref{BHs}, we first give a brief review of a LQG black hole and then analyze the deflection angle of light rays by such a black hole. In Secs.~\ref{wdl} and \ref{sdl}, we calculate the bending angles of light rays in the weak and strong deflection limits, respectively. The observables of the relativistic images are investigated in Sec.~\ref{osdl}. Section \ref{con} comes with the conclusion.

\section{Black holes in effective loop quantum gravity}~\label{BHs}

Let us start with the quantum extension of the static and spherically symmetric metric by solving the loop quantum gravity (LQG) effective equations \cite{Bodendorfer136390,Bodendorfer095002,Brahma181301}: 
\begin{eqnarray}
ds^2=-\mathcal{A}(y)d\tau^2+\frac{dy^2}{\mathcal{A}(y)}+\mathcal{C}^2(y)d\Omega_2^2, ~\label{lqgbh}
\end{eqnarray}
where the metric functions in terms of the radial coordinate $y\in (-\infty,\infty)$ are given by
\begin{eqnarray}
\mathcal{A}(y)&=&\left(1-\frac{1}{\sqrt{2A_{\lambda}(1+y^2)}}\right)\frac{1+y^2}{\mathcal{C}^2(y)}, \\
\mathcal{C}^2(y)&=&\frac{A_{\lambda}}{\sqrt{1+y^2}}\frac{M_B^2(y+\sqrt{1+y^2})^6+M_W^2}{(y+\sqrt{1+y^2})^3},
\end{eqnarray}
with $M_B$ and $M_W$ two Dirac observables in the model. The dimensionless and non-negative parameter $A_{\lambda}$ is defined as $A_{\lambda}\equiv (\lambda_k/(M_B M_W))^{2/3}/2$, where $\lambda_k$ is a quantum parameter originated from holonomy modifications \cite{Bodendorfer136390,Bodendorfer095002}. An important feature of this black hole is that inside the black hole, the areal radius $\mathcal{C}$ reaches a minimum, which stands for a spacelike transition surface that smoothly connects an asymptotically Schwarzschild black hole to a white hole with mass $M_B$ and $M_W$, respectively \cite{Brahma181301}. In this paper, we focus on the physically interesting case with $M_B=M_W=M$ corresponding to the symmetric bounce, i.e., the spacetime is symmetric under $y\rightarrow -y$. Without loss of generality, we concentrate on the branch of $y\geqslant 0$. After redefining two new coordinates $r\equiv\sqrt{8A_{\lambda}}M_By$ and $t\equiv\tau/(\sqrt{8A_{\lambda}}M_B)$, the metric (\ref{lqgbh}) can be rewritten as \cite{Brahma181301}
\begin{eqnarray}
ds^2=-A(r)dt^2+B(r)dr^2+C(r)(d\theta^2+\sin^2\theta d\phi^2), ~\label{metric1}
\end{eqnarray}
where
\begin{eqnarray}
A(r)\!\!&=&\!\!\frac{1}{B(r)}\!=\!\frac{\sqrt{8A_{\lambda}M^2+r^2}\big(\sqrt{8A_{\lambda}M^2+r^2}-2M\big)}{2A_{\lambda}M^2+r^2}, ~~~\\
C(r)\!\!&=&\!\!2A_{\lambda}M^2+r^2.
\end{eqnarray}
{Then, the radius of the horizon is $r_h=2M\sqrt{1-2A_{\lambda}}$. Near the horizon $r=r_h+\epsilon$, $A(r)$ can be expanded as
\begin{eqnarray}
A(r_h+\epsilon)=\frac{\sqrt{1-2A_{\lambda}}\epsilon}{(2-3A_{\lambda})M}+\mathcal{O}(\epsilon^2).
\end{eqnarray}
Obviously, if $A_{\lambda}\ll\frac{1}{2}$, $A(r)\simeq\frac{\epsilon}{2M}$ near the horizon, which is the same as the Schwarzschild case. Besides, when $r\rightarrow \infty$, we have $A(r)\rightarrow 1-\frac{2M}{r}$. Namely, the spacetime reduces to the Schwarzschild black hole when it approaches infinity. Thus, the metric (\ref{metric1}) can not be distinguishable from the Schwarzschild black hole at or beyond the horizon for $A_{\lambda}\ll\frac{1}{2}$.
}

It can be easily shown that the metric (\ref{metric1}) represents (i) a Schwarzschild black hole if $A_{\lambda}=0$ and $M\neq 0$, (ii) a regular black hole if $0<A_{\lambda}<\frac{1}{2}$, (iii) a one-way traversable wormhole with a null throat if $A_{\lambda}=\frac{1}{2}$, (vi) a traversable wormhole with a two-way throat at $r=0$ if $A_{\lambda}>\frac{1}{2}$, and (v) a flat spacetime if $M=0$. From the line element (\ref{metric1}), the Lagrangian for the geodesics can be expressed as
\begin{eqnarray}
2\mathcal{L}=-A(r)\dot{t}^2+B(r)\dot{r}^2+C(r)(\dot{\theta}^2+\sin^2\theta\dot{\phi}^2),
\end{eqnarray}
where the dot denotes the derivative with respect to an affine parameter along the trajectory. Since the independence of the Lagrangian on $t$ and $\phi$, one can obtain two conserved quantities:
\begin{eqnarray}
p_t&=&\frac{\partial\mathcal{L}}{\partial\dot{t}}=-A(r)\dot{t}=-E, \\
p_{\phi}&=&\frac{\partial\mathcal{L}}{\partial\dot{\phi}}=C(r)\dot{\phi}\sin^2\theta.
\end{eqnarray}

\begin{figure}[htb]
\begin{center}
\includegraphics[width=8cm]{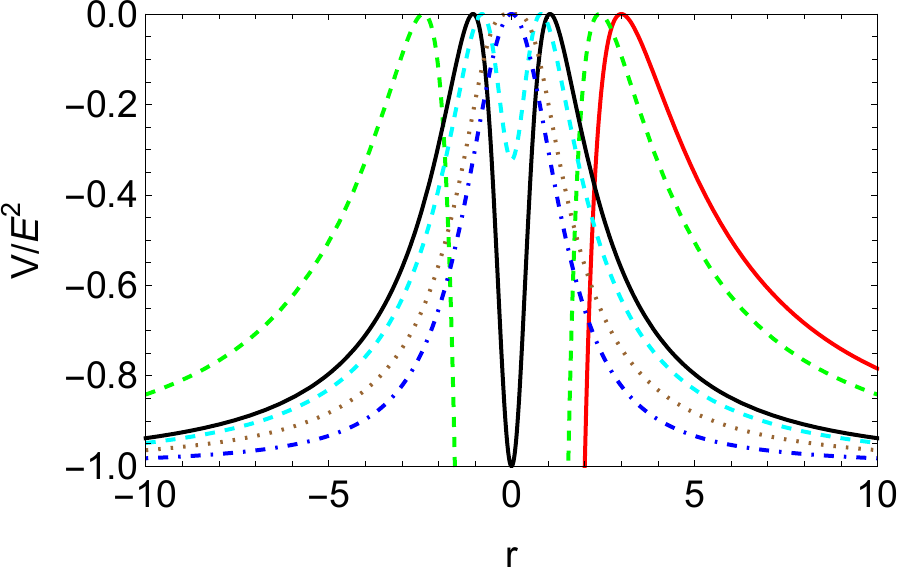}
\end{center}
\caption{Dimensionless effective potential $V/E^2$ in terms of the radial coordinate $r$. The parameters are set to $M=1$, $A_{\lambda}=0$ for red line (Schwarzschild metric), $A_{\lambda}=0.2$ for green dashed line (regular black hole), $A_{\lambda}=\frac{1}{2}$ for black line (one-way traversable wormhole), $A_{\lambda}=0.53$ for cyan dashed line (traversable wormhole ), $A_{\lambda}=\frac{225}{392}$ for brown dotted line (traversable wormhole with a marginally unstable photon sphere), and $A_{\lambda}=2$ for blue dotted dashed line (traversable wormhole).}
\label{Vr}
\end{figure}

The lightlike geodesics in the equatorial plane $\theta=\frac{\pi}{2}$ are given by
\begin{eqnarray}
-A(r)\dot{t}^2+B(r)\dot{r}^2+C(r)\dot{\phi}^2=0. ~\label{geoeq}
\end{eqnarray}
Inserting the two conserved quantities into Eq.~(\ref{geoeq}), the null geodesics can be cast as
\begin{eqnarray}
\dot{r}^2+V(r)=0,
\end{eqnarray}
where the effective potential is $V(r)\equiv\frac{L^2}{C(r)B(r)}-E^2$ and $L\equiv C(r)\dot{\phi}$.
Obviously, light rays can only exist in a region for $V(r)\leq 0$. The effective potential and its derivatives with respect to the radial coordinate $r$ can be calculated as
\begin{widetext}
\begin{eqnarray}
V(r)\!\!&=&\!\!E^2\left(b^2\frac{\left(8 M^2 A_{\lambda }-2 M \sqrt{8 M^2 A_{\lambda }+r^2}+r^2\right)}{\left(2 M^2 A_{\lambda }+r^2\right){}^2}-1\right), \\
V'(r)\!\!&=&\!\!-\frac{2b^2E^2 r}{\left(2 A_{\lambda} M^2+r^2\right)^3 \sqrt{8 A_{\lambda} M^2+r^2}}\bigg(2 A_{\lambda} M^2 \left(7 \sqrt{8 A_{\lambda} M^2+r^2}-15 M\right)+r^2 \left(\sqrt{8 A_{\lambda} M^2+r^2}-3 M\right)\bigg), \\
V''(r)\!\!&=&\!\!\frac{2b^2E^2}{\left(2 A_{\lambda} M^2+r^2\right)^4 \left(8 A_{\lambda} M^2+r^2\right)^{3/2}}\bigg(32 A_{\lambda}^3 M^6 \left(15 M-7 \sqrt{8 A_{\lambda} M^2+r^2}\right)+44 A_{\lambda}^2 M^4 r^2 \bigg(11 \sqrt{8 A_{\lambda} M^2+r^2} \nonumber\\
       \!\!&-&\!\!24 M\bigg)+3 r^6 \left(\sqrt{8 A_{\lambda} M^2+r^2}-4 M\right)+8 A_{\lambda} M^2 r^4 \left(11 \sqrt{8 A_{\lambda} M^2+r^2}-30 M\right)\bigg), \\
V'''(r)\!\!&=&\!\!-\frac{24b^2E^2 r}{\left(2 A_{\lambda} M^2+r^2\right)^5 \left(8 A_{\lambda} M^2+r^2\right)^{5/2}}\bigg(16 A_{\lambda}^4 M^8 \left(351 M-160 \sqrt{8 A_{\lambda} M^2+r^2}\right)+40 A_{\lambda}^3 M^6 r^2 \big(32 \sqrt{8 A_{\lambda} M^2+r^2} \nonumber\\
       \!\!&-&\!\!67 M\big)+72 A_{\lambda}^2 M^4 r^4 \left(7 \sqrt{8 A_{\lambda} M^2+r^2}-18 M\right)+r^8 \left(\sqrt{8 A_{\lambda} M^2+r^2}-5 M\right)+2 A_{\lambda} M^2 r^6 \big(23 \sqrt{8 A_{\lambda} M^2+r^2} \nonumber\\
       \!\!&-&\!\!75 M\big)\bigg), \\
V''''(r)\!\!&=&\!\!\frac{24b^2E^2}{\left(2 M^2 A_{\lambda }+r^2\right){}^6 \left(8 M^2 A_{\lambda }+r^2\right){}^{7/2}}\bigg(5 r^{12} \left(\sqrt{8 M^2 A_{\lambda }+r^2}-6 M\right)+80 M^2 r^{10} A_{\lambda } \left(4 \sqrt{8 M^2 A_{\lambda }+r^2}-15 M\right) \nonumber\\
        \!\!&+&\!\!256 M^{12} A_{\lambda }^6 \left(160 \sqrt{8 M^2 A_{\lambda }+r^2}-351 M\right)+1920 M^{10} r^2 A_{\lambda }^5 \left(301 M-136 \sqrt{8 M^2 A_{\lambda }+r^2}\right) \nonumber\\
        \!\!&+&\!\!128 M^8 r^4 A_{\lambda }^4 \left(5 \sqrt{8 M^2 A_{\lambda }+r^2}+168 M\right)+240 M^6 r^6 A_{\lambda }^3 \left(117 \sqrt{8 M^2 A_{\lambda }+r^2}-280 M\right) \nonumber\\
        \!\!&+&\!\!60 M^4 r^8 A_{\lambda }^2 \left(87 \sqrt{8 M^2 A_{\lambda }+r^2}-254 M\right)\bigg),
\end{eqnarray}
\end{widetext}
where the impact parameter is defined as $b(r_0)\equiv\frac{L}{E}=\sqrt{\frac{C(r_0)}{A(r_0)}}$, and $r_0$ is the closest distance between the light ray and the black hole. Light rays with the critical impact parameter $b_m\equiv b(r_m)$ form unstable circular orbits called photon sphere, where $r_m$ is the radius of the photon sphere, which is determined by $V(r_m)=V'(r_m)=0$ and $V''(r_m)\leqslant0$. Light rays with the impact parameter $b<b_m$ are captured by the black hole or the wormhole while light rays with $b>b_m$ are deflected. The dimensionless effective potential $V(r)/E^2$ for the light ray with the critical impact parameter $b_m$ is plotted in Fig.~\ref{Vr}. Obviously,  for $0< A_{\lambda}<\frac{225}{392}$,  there are two unstable photon spheres located at $r=\pm r_m$, respectively. For $A_{\lambda}\geqslant\frac{225}{392}$, these two unstable photon spheres merge into a marginally unstable photon sphere located at $r_m=0$. Besides, it should be noted that the light orbits at $r=0$ satisfy $V'(0)=0$ and $V''(0)>0$ for $\frac{1}{2}\leqslant A_{\lambda}<\frac{225}{392}$, however $V(0)\neq 0$, which indicates $\dot{r}\neq 0$. 
Thus there is no antiphoton sphere at $r=0$ according to its definition in Refs.~\cite{Tsukamoto024033,Tsukamoto064022}.

From Eq.~(\ref{geoeq}), the deflection angle $\alpha(r_0)$ can be computed to be
\begin{eqnarray}
\alpha(r_0)=I(r_0)-\pi,~\label{exdeflection}
\end{eqnarray}
where
\begin{eqnarray}
I(r_0)\equiv 2\int^{\infty}_{r_0}\frac{dr}{\sqrt{\frac{C}{B}\left(\frac{C}{Ab^2}-1\right)}}.~\label{Ir0}
\end{eqnarray}

\section{Weak deflection angle} ~\label{wdl}

In this section, we investigate the weak deflection angle of light rays in this LQG black hole spacetime. By defining $x=r/2M$ and $T=t/2M$, the line element (\ref{metric1}) can be transformed into
\begin{eqnarray}
dS^2&=&(2M)^{-2}ds^2=-A(x)dT^2+B(x)dx^2 \nonumber\\
        &+&C(x)\big(d\theta^2+\sin^2\theta d\phi^2\big),~\label{dS1}
\end{eqnarray}
where
\begin{eqnarray}
A(x)&=&\frac{1}{B(x)}=\frac{2\sqrt{2A_{\lambda}+x^2}(\sqrt{2A_{\lambda}+x^2}-1)}{A_{\lambda}+2x^2},~\label{Ax} \\
C(x)&=&\frac{A_{\lambda}}{2}+x^2.~\label{Cx}
\end{eqnarray}
Then, the integral (\ref{Ir0}) reduces to
\begin{eqnarray}
I(x_0)&=&2\int^{\infty}_{x_0}(A_{\lambda }+2 x^2)^{-1}\bigg(\frac{-2 A_{\lambda }+\sqrt{2 A_{\lambda }+x^2}-x^2}{\left(A_{\lambda }+2 x^2\right){}^2} \nonumber\\
         &+&\frac{2 A_{\lambda }-\sqrt{2 A_{\lambda }+x_0^2}+x_0^2}{\left(A_{\lambda }+2 x_0^2\right){}^2}\bigg)^{-1/2}dx, ~\label{Ix0}
\end{eqnarray}
and the impact parameter is given by
\begin{eqnarray}
\frac{b}{2M}=\left(\frac{A_{\lambda}}{2}+x_0^2\right)\bigg(2A_{\lambda}+x_0^2-\sqrt{2A_{\lambda}+x_0^2}\bigg)^{-1/2}.~\label{bb}
\end{eqnarray}

After introducing a new variable $z=x_0/x$, the above integral can be rewritten as
\begin{eqnarray}
I(x_0)\!\!&=&\!\!\int^1_0\frac{2x_0}{z^2}\left(A_{\lambda }+\frac{2 x_0^2}{z^2}\right)^{-1}\!\!\bigg(\frac{2 A_{\lambda }\!-\!\sqrt{2 A_{\lambda }+x_0^2}\!+\!x_0^2}{\left(A_{\lambda }+2 x_0^2\right){}^2} \nonumber\\
                       &+&\frac{-2 A_{\lambda }z^2+z\sqrt{2 A_{\lambda }z^2+x_0^2}-x_0^2}{z\left(A_{\lambda }z^2+2 x_0^2\right){}^2}\bigg)^{-1/2}dz,~\label{wIx0}
\end{eqnarray}
which can not be integrated out analytically. However, in the weak field limit, i.e., $x_0\gg 1$, the integrand can be expanded in terms of $1/x_0$, which can be integrated out term by term as follows:
\begin{eqnarray}
\alpha(x_0)&=&I(x_0)-\pi=\frac{2}{x_0}+\left(\pi  \left(\frac{15}{16}-A_{\lambda }\right)-1\right)\frac{1}{x_0^2} \nonumber\\
                  &+&\left(\left(\pi -\frac{25}{3}\right) A_{\lambda }+\frac{61}{12}-\frac{15 \pi }{16}\right)\frac{1}{x_0^3} \nonumber\\
                  &+&\bigg(\frac{1}{64} A_{\lambda } \big(152 \pi  A_{\lambda }-465 \pi +928\big)+\frac{3465 \pi }{1024} \nonumber\\
                  &-&\frac{65}{8}\bigg)\frac{1}{x_0^4}+\bigg(-\frac{1}{480} A_{\lambda } \big(24 (135 \pi -746) A_{\lambda } \nonumber\\
                  &-&7875 \pi +30940\big)+\frac{7783}{320}-\frac{3465 \pi }{512}\bigg)\frac{1}{x_0^5} \nonumber\\
                  &+&\bigg(\frac{1}{512} A_{\lambda } \big(-3064 \pi  A_{\lambda }^2+3 (8601 \pi -21376) A_{\lambda} \nonumber\\
                  &-&30765 \pi +86560\big)+\frac{310695 \pi }{16384}-\frac{21397}{384}\bigg)\frac{1}{x_0^6} \nonumber\\
                  &+&\mathcal{O}\left(\frac{1}{x_0^7}\right).
\end{eqnarray}
Besides, the inverse solution of Eq.~(\ref{bb}) also can be obtained in the weak field limit as
\clearpage
\begin{eqnarray}
\frac{1}{x_0}&=&\frac{2M}{b}+\frac{1}{2}\left(\frac{2M}{b}\right)^2+\frac{1}{8}(5-4A_{\lambda})\left(\frac{2M}{b}\right)^3 \nonumber\\
                   &+&\left(1-\frac{3A_{\lambda}}{2}\right)\left(\frac{2M}{b}\right)^4+\frac{3}{128} \big(64 A_{\lambda }^2-168 A_{\lambda} \nonumber\\
                   &+&77\big)\left(\frac{2M}{b}\right)^5
                   +\left(\frac{57 A_{\lambda }^2}{8}-10 A_{\lambda }+\frac{7}{2}\right)\left(\frac{2M}{b}\right)^6 \nonumber\\
                   &+&\mathcal{O}\left(\left(\frac{2M}{b}\right)^7\right).
\end{eqnarray}

Inserting the above relation into Eq.~(\ref{wIx0}), the weak deflection angle as a function of $b$ is found to be
\begin{eqnarray}
\alpha&=&\frac{4M}{b}+\frac{15\pi M^2}{4b^2}-\frac{4\pi A_{\lambda}M^2}{b^2}+\frac{(128-224A_{\lambda})M^3}{3b^3} \nonumber\\
          &+&\mathcal{O}(b^{-4}).~\label{wedeflection}
\end{eqnarray}
It is obvious that the first two terms come from the weak deflection angle of light rays by a spherically symmetric black hole \cite{Gibbons235009}, while the third term originates from the quantum effects, which is of the same order as the charge or spin of a black hole \cite{Werner3047,Jusufi24}. Besides, the minus sign in front of $A_{\lambda}$ indicates that the quantum effects make a negative contribution to the weak deflection angle. {What' s more, the primary quantum correction to the weak deflection angle from the LQG is larger than the non-loop quantum one investigated in Ref.~\cite{Bjerrum-Bohr061301}, where the primary correction is only at the order of $b^{-3}$.}

\section{Strong deflection limit} ~\label{sdl}

In this section, we mainly focus on the bending angle in the strong deflection limit, i.e., $x_0\rightarrow x_m$ or $b\rightarrow b_m$, for different values of $A_{\lambda}$.

\subsection{Case of $0\leqslant A_{\lambda}<\frac{225}{392}$}

For $0\leqslant A_{\lambda}<\frac{225}{392}$, there are two unstable photon spheres at $x=\pm x_m$, respectively. Without loss of generality, we focus on the unstable photon sphere at $x=x_m$ and investigate the deflection angle of light rays in the strong deflection limit. After redefining a new variable $\bar{z}$ as 
\begin{eqnarray}
\bar{z}=1-\frac{x_0}{x}, ~\label{zc}
\end{eqnarray}
the integral (\ref{Ir0}) can be rewritten as
\begin{eqnarray}
I(x_0)=\int^1_0 R(\bar{z},x_0)f(\bar{z},x_0)d\bar{z},
\end{eqnarray}
where
\begin{eqnarray}
R(\bar{z},x_0)&=&\frac{2x_0\sqrt{ABC_0}}{C(1-\bar{z})^2}, \\
f(\bar{z},x_0)&=&\frac{1}{\sqrt{A_0-\frac{C_0}{C}A}}.
\end{eqnarray}
It can be easily shown that the function $R(\bar{z},x_0)$ is regular for all values of $\bar{z}$ and $x_0$, while $f(\bar{z},x_0)$ diverges when $\bar{z}\rightarrow 0$. Then, following Ref.~\cite{Bozza103001}, the integral $I(x_0)$ can be divided into two parts:
\begin{eqnarray}
I(x_0)=I_D(x_0)+I_R(x_0),
\end{eqnarray}
i.e., the divergent part $I_D(x_0)$ and the regular part $I_R(x_0)$. The divergent part including the divergence is given by
\begin{eqnarray}
I_D(x_0)=\int^1_0 R(0,x_m)f_D(\bar{z},x_0)d\bar{z},
\end{eqnarray}
where $f_D(\bar{z},x_0)$ is the series expansion of the argument of the square root in $f(\bar{z},x_0)$ up to the second order in $\bar{z}$:
\begin{eqnarray}
f_D(\bar{z},x_0)=\frac{1}{\sqrt{c_1(x_0)\bar{z}+c_2(x_0)\bar{z}^2}},
\end{eqnarray}
with the expansion coefficients $c_1$ and $c_2$ given by
\begin{eqnarray}
c_1(x_0)&=&x_0\left(\frac{A_0C_0'}{C_0}-A_0'\right), \\
c_2(x_0)&=&\frac{x_0}{2C_0^2}\big(-2x_0A_0C_0'^2-C_0^2(2A_0'+x_0A_0'') \nonumber\\
              &+&C_0(2x_0A_0'C_0'+A_0(2C_0'+x_0C_0''))\big),
\end{eqnarray}
where the functions with subscript $0$ denote that they are computed at $x_0$.

After substracting the divergent part, one can obtain the regular part:
\begin{eqnarray}
I_R(x_0)=\int^1_0 g_R(\bar{z},x_0)d\bar{z},~\label{IR}
\end{eqnarray}
where $g_R(\bar{z},x_0)\equiv R(\bar{z},x_0)f(\bar{z},x_0)-R(0,x_m)f_D(\bar{z},x_0)$. Then, the deflection angle of light rays in the strong field limit can be expressed as
\begin{eqnarray}
\alpha(b)\!=\!-\bar{a}\ln\left(\frac{b}{b_m}-1\right)\!+\!\bar{u}\!+\!\mathcal{O}[(b-b_m)\ln(b-b_m)], ~\label{deflection}
\end{eqnarray}
where
\begin{eqnarray}
\bar{a}&\equiv&\frac{R(0,x_m)}{2\sqrt{c_2(x_m)}},~\label{abar} \\
\bar{u}&\equiv&\bar{a}\ln\bar{\delta}+I_R(x_0)-\pi, \\
\bar{\delta}&\equiv&x_m^2\left(\frac{C''(x_m)}{C(x_m)}-\frac{A''(x_m)}{A(x_m)}\right).~\label{dbar}
\end{eqnarray}
After inserting Eqs.~(\ref{Ax}) and (\ref{Cx}) into Eqs.~(\ref{abar}) and (\ref{dbar}), the coefficients in the deflection angle (\ref{deflection}) can be calculated as
\begin{widetext}
\begin{eqnarray}
\bar{a}(x_m)&=&\frac{\left(2 A_{\lambda }+x_m^2\right){}^{3/4} \left(A_{\lambda }+2 x_m^2\right)}{\sqrt{4 x_m^6 \sqrt{2 A_{\lambda }+x_m^2}+12 A_{\lambda } x_m^4+3 A_{\lambda }^2 x_m^2 \left(4-3 \sqrt{2 A_{\lambda }+x_m^2}\right)+A_{\lambda }^3 \left(14 \sqrt{2 A_{\lambda }+x_m^2}-15\right)}}, \\
\bar{\delta}(x_m)&=&-\frac{2 x_m^2 \left(4 x_m^6 \sqrt{2 A_{\lambda }+x_m^2}+12 A_{\lambda } x_m^4+3 A_{\lambda }^2 x_m^2 \left(4-3 \sqrt{2 A_{\lambda }+x_m^2}\right)+A_{\lambda }^3 \left(14 \sqrt{2 A_{\lambda }+x_m^2}-15\right)\right)}{\left(2 A_{\lambda }+x_m^2\right){}^{3/2} \left(A_{\lambda }+2 x_m^2\right){}^2 \left(-2 A_{\lambda }+\sqrt{2 A_{\lambda }+x_m^2}-x_m^2\right)}.
\end{eqnarray}
\end{widetext}

\begin{figure*}[htb]
\begin{center}
\subfigure  {\label{aa1}
\includegraphics[width=5.3cm]{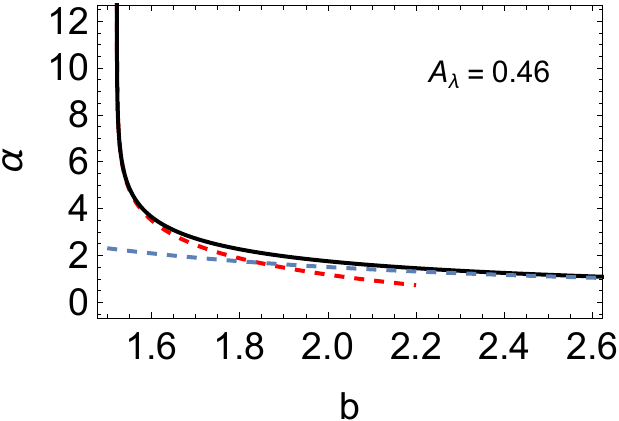}}
\subfigure  {\label{aa2}
\includegraphics[width=5cm]{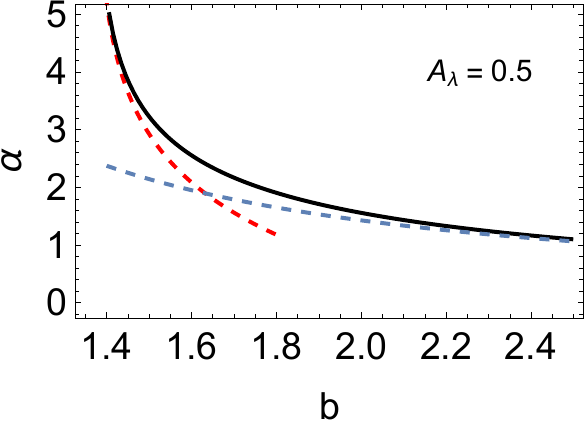}}
\subfigure  {\label{aa3}
\includegraphics[width=5cm]{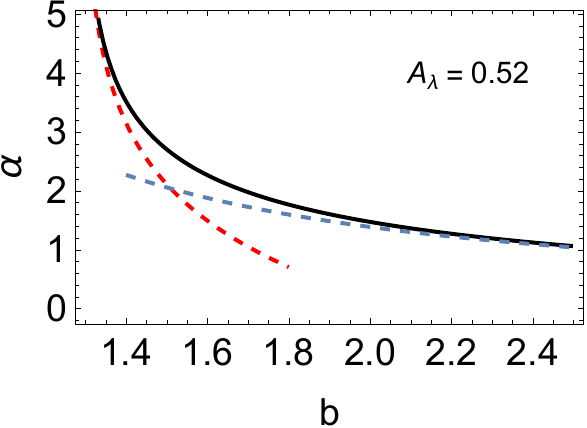}}
\end{center}
\caption{Plots of the exact deflection angle (\ref{exdeflection}) (the black full lines), the corresponding strong deflection limit (\ref{deflection}) (the red dashed lines) and weak deflection limit (\ref{wedeflection}) (the blue dashed lines) as a function of the impact parameter $b$ for some values of $A_{\lambda}$.}
\label{AA1}
\end{figure*}

Expanding the integrand of the regular part (\ref{IR}) in powers of $x_0-x_m$:
\begin{eqnarray}
I_R(x_0)=\sum^{\infty}_{j=0}\frac{1}{j!}(x_0-x_m)^j \int^1_0 \frac{\partial^j g}{\partial x^j_0}\Bigg |_{x_0=x_m}d\bar{z},
\end{eqnarray}
and focusing on the primary term $j=0$, the regular part can be approximated as
\begin{eqnarray}
I_R(x_0)\sim I_R(x_m)=\int^1_0 g(\bar{z},x_m)d\bar{z},
\end{eqnarray}
which can be calculated numerically. Figure \ref{AA1} shows the deflection angle of light rays (\ref{deflection}) in terms of $b$ for several values of $A_{\lambda}$. It can be seen that the strong deflection angle is an excellent approximation for photons passing close to the photon sphere, as shown in Fig.~\ref{AA1}. The weak deflection angle (\ref{wedeflection}) is also a good approximation to the exact deflection angle (\ref{exdeflection}). Besides, for $A_{\lambda}=0$, the strong deflection angle of light rays by Schwarzschild black hole is recovered \cite{Bozza103001}:
\begin{eqnarray}
\alpha(b)&=&-\ln\left(\frac{b}{3\sqrt{3}M}-1\right)+\ln6+0.9496-\pi \nonumber\\
              &+&\mathcal{O}[(b-b_m)\ln(b-b_m)].
\end{eqnarray}

\subsection{Case of $A_{\alpha}>\frac{225}{392}$}

In the case of $A_{\alpha}>\frac{225}{392}$, two unstable photon spheres merge into one unstable photon sphere located at  the wormhole throat $x=0$. Defining a new radial coordinate $\rho=\sqrt{\frac{A_{\lambda}}{2}+x^2}$, the line element (\ref{dS1}) can be transformed into
\begin{eqnarray}
dS^2\!=\!-\tilde{A}(\rho)dT^2\!+\!\tilde{B}(\rho)d\rho^2\!+\!\tilde{C}(\rho)(d\theta^2\!+\!\sin^2\theta d\phi^2),
\end{eqnarray}
where
\begin{eqnarray}
\tilde{A}(\rho)&=&1+\frac{3 A_{\lambda }}{2 \rho ^2}-\frac{\sqrt{6 A_{\lambda }+4 \rho ^2}}{2 \rho ^2}, \\
\tilde{B}(\rho)&=&\frac{2\rho^2}{(2\rho^2-A_{\lambda})\tilde{A}(\rho)}, \\
\tilde{C}(\rho)&=&\rho^2.
\end{eqnarray}

Since the variable $\bar{z}$ defined in Eq.~(\ref{zc}) would not work well in this case, we introduce a new definition $\tilde{z}=1-\frac{\rho_0}{\rho}$. Then, the integral (\ref{Ir0}) can be expressed as
\begin{eqnarray}
I(\rho_0)=\int^1_0 \tilde{R}(\tilde{z},\rho_0)\tilde{f}(\tilde{z},\rho_0)d\tilde{z},~\label{Irho}
\end{eqnarray}
where
\begin{eqnarray}
\tilde{R}(\tilde{z},\rho_0)&=&\frac{\rho_0\sqrt{\tilde{C}(\rho_0)}}{\tilde{C}(\rho)(1-\tilde{z})^2}, \\
\tilde{f}(\tilde{z},\rho_0)&=&\sqrt{\frac{\tilde{A}(\rho)\tilde{B}(\rho)}{\tilde{A}(\rho_0)-\frac{\tilde{C}(\rho_0)}{\tilde{C}(\rho)}\tilde{A}(\rho)}},
\end{eqnarray}
and $\rho_0=\sqrt{\frac{A_{\lambda}}{2}+x_0^2}$.
After inserting $\tilde{A}(\rho)$, $\tilde{B}(\rho)$ and $\tilde{C}(\rho)$, the above expressions can be simplified to
\begin{eqnarray}
\tilde{R}(\tilde{z},\rho_0)&=&2, \\
\tilde{f}(\tilde{z},\rho_0)&=&\frac{2\rho_0^2}{\sqrt{\tilde{G}(\tilde{z},\rho_0)}},~\label{ft}
\end{eqnarray}
where
\begin{eqnarray}
\tilde{G}(\tilde{z},\rho_0)&=&\left(2 \rho _0^2-(\tilde{z}-1)^2 A_{\lambda }\right) \bigg[2 \rho _0^2-\sqrt{6 A_{\lambda }+4 \rho _0^2} \nonumber\\
                                      &+&3 A_{\lambda }-\Big((\tilde{z}-1)^2 \big((\tilde{z}-1) \sqrt{6 (\tilde{z}-1)^2 A_{\lambda }+4 \rho _0^2} \nonumber\\
                            &+&3 (\tilde{z}-1)^2 A_{\lambda }+2 \rho _0^2\big)\Big)\bigg].
\end{eqnarray}
Obviously, $\tilde{R}(\tilde{z},\rho_0)$ is regular and $\tilde{f}(\tilde{z},\rho_0)$ is divergent when $\tilde{z}\rightarrow 0$. To solve the integral (\ref{Irho}), we also split it into two parts $I(\rho_0)=I_d(\rho_0)+I_r(\rho_0)$. The divergent part $I_d(\rho_0)$ is given by
\begin{eqnarray}
I_d(\rho_0)&=&\int^1_0 2\tilde{f}_d(\tilde{z},\rho_0)d\tilde{z}, \\
\tilde{f}_d(\tilde{z},\rho_0)&=&\frac{2\rho_0^2}{\sqrt{c_3(\rho_0)\tilde{z}+c_4(\rho_0)\tilde{z}^2+c_5(\rho_0)\tilde{z}^3}},~\label{fd}
\end{eqnarray}
where $\rho_m=\sqrt{\frac{A_{\lambda}}{2}+x_m^2}$, and the expansion coefficients of the function $\tilde{G}(\tilde{z},\rho_0)$ up to the third order in $\tilde{z}$ are given by
\begin{eqnarray}
c_3(\rho_0)\!\!&=&\!\!2 \rho _0^3 \left(A_{\lambda }-2 \rho _0^2\right) \left(\tilde{A}_0'-\frac{\tilde{A}_0 \tilde{C}_0'}{\tilde{C}_0}\right), \\
c_4(\rho_0)\!\!&=&\!\!-\frac{\rho _0^3}{\tilde{C}_0^2}\bigg(\tilde{C}_0^2 \left(2 \tilde{A}_0' \left(A_{\lambda }+2 \rho _0^2\right)\!+\!\rho _0 \tilde{A}_0'' \left(2 \rho _0^2-A_{\lambda }\right)\right) \nonumber\\
                  \!\!&+&\!\!\tilde{C}_0 \Big(\tilde{A}_0 \left(\rho _0 \tilde{C}_0'' \left(A_{\lambda }-2 \rho _0^2\right)-2 \tilde{C}_0' \left(A_{\lambda }+2 \rho _0^2\right)\right) \nonumber\\
                  \!\!&+&\!\!2 \rho _0 \tilde{A}_0' \tilde{C}_0' \left(A_{\lambda }-2 \rho _0^2\right)\Big) \nonumber\\
                  \!\!&+&\!\!2 \rho _0 \tilde{A}_0 \tilde{C}_0'^2 \left(2 \rho _0^2-A_{\lambda }\right)\bigg), \\
c_5(\rho_0)\!\!&=&\!\!-\frac{\rho _0^5 }{3 \tilde{C}_0^3}\bigg(\tilde{C}_0^3 \big(-\tilde{A}_0{}^{(3)} \left(A_{\lambda }-2 \rho _0^2\right)+12 \rho _0 \tilde{A}_0'' \nonumber\\
                  \!\!&+&\!\!12 \tilde{A}_0'\big)-\tilde{C}_0^2 \Big(-3 \tilde{A}_0'' \tilde{C}_0' \left(A_{\lambda }-2 \rho _0^2\right) \nonumber\\
                  \!\!&+&\!\!3 \tilde{A}_0' \left(8 \rho _0 \tilde{C}_0'-\tilde{C}_0'' \left(A_{\lambda }-2 \rho _0^2\right)\right) \nonumber\\
                  \!\!&+&\!\!\tilde{A}_0 \left(-\tilde{C}_0{}^{(3)} \left(A_{\lambda }-2 \rho _0^2\right)+12 \rho _0 \tilde{C}_0''+12 \tilde{C}_0'\right)\Big) \nonumber\\
                  \!\!&+&\!\!6 \tilde{C}_0 \tilde{C}_0' \Big(\tilde{A}_0 \left(4 \rho _0 \tilde{C}_0'-\tilde{C}_0'' \left(A_{\lambda }-2 \rho _0^2\right)\right) \nonumber\\
                  \!\!&-&\!\!\tilde{A}_0' \tilde{C}_0' \left(A_{\lambda }-2 \rho _0^2\right)\Big)+6 \tilde{A}_0 \tilde{C}_0'^3 \left(A_{\lambda }-2 \rho _0^2\right)\bigg),
\end{eqnarray}
where the functions with subscript $0$ denote that they are computed at $\rho_0$.

Expanding $c_3(\rho_0)$, $c_4(\rho_0)$, and $c_5(\rho_0)$ in powers of $\rho_0-\rho_m$:
\begin{eqnarray}
c_3(\rho_0)\!&=&\!\left(28 \sqrt{2} A_{\lambda }^{3/2}-30 A_{\lambda }\right)(\rho_0-\rho_m)+\bigg(44 A_{\lambda } \nonumber\\
                   \!&-&\!\frac{39 \sqrt{A_{\lambda }}}{\sqrt{2}}\bigg)(\rho_0-\rho_m)^2+\mathcal{O}\big((\rho_0-\rho_m)^3\big), ~\label{c3} \\
c_4(\rho_0)\!&=&\!\left(28 A_{\lambda }^2-15 \sqrt{2} A_{\lambda }^{3/2}\right)+\bigg(\frac{153 A_{\lambda }}{4} \nonumber\\
                  \!&-&\!30 \sqrt{2} A_{\lambda }^{3/2}\bigg)(\rho_0-\rho_m)+\mathcal{O}\big((\rho_0-\rho_m)^2\big), \label{c4} \\
c_5(\rho_0)\!&=&\!\left(\frac{231 A_{\lambda }^{3/2}}{4 \sqrt{2}}-52 A_{\lambda }^2\right)+\mathcal{O}\big((\rho_0-\rho_m)\big), ~\label{c5}
\end{eqnarray}
one can easily see that $\tilde{f}_d(\tilde{z},\rho_0)$ diverges as $\tilde{z}^{-1}$ in the strong deflection limit $\rho_0\rightarrow \rho_m$.

Subtracting the divergent part from $I(\rho_0)$, the regular part can be obtained as
\begin{eqnarray}
I_r(\rho_0)=\int^1_0 \tilde{g}(\tilde{z},\rho_0)d\tilde{z},
\end{eqnarray}
where $\tilde{g}(\tilde{z},\rho_0)\equiv2\left(\tilde{f}(\tilde{z},\rho_0)-\tilde{f}_d(\tilde{z},\rho_0)\right)$.
Then, the deflection angle of light rays in the strong deflection limit $\rho_0\rightarrow \rho_m$ is
\begin{eqnarray}
\alpha(b)\!=\!-\tilde{a}\ln\left(\frac{b}{b_m}-1\right)\!+\!\tilde{u}\!+\!\mathcal{O}[(b-b_m)\ln(b-b_m)],~\label{deflection2}
\end{eqnarray}
where
\begin{eqnarray}
\tilde{a}&\equiv&\frac{4\rho_m^2}{\sqrt{c_4(\rho_m)}}, \\
\tilde{u}&\equiv&\tilde{a}\ln\tilde{\delta}+I_r(\rho_0)-\pi, \\
\tilde{\delta}&\equiv&2\rho_m\left(\frac{\tilde{C}'(\rho_m)}{\tilde{C}(\rho_m)}-\frac{\tilde{A}'(\rho_m)}{\tilde{A}(\rho_m)}\right).
\end{eqnarray}
By inserting $\tilde{A}(\rho)$, $\tilde{B}(\rho)$ and $\tilde{C}(\rho)$, the parameters $\tilde{a}$ and $\tilde{\delta}$ reduce to
\begin{eqnarray}
\tilde{a}(\rho_m)\!&=&\!\rho_m\sqrt{6A_{\lambda}+4\rho_m^2}\bigg(18 A_{\lambda }^2-3 \rho _m^2 \sqrt{6 A_{\lambda }+4 \rho _m^2} \nonumber\\
                          \!&-&\!6 A_{\lambda } \left(\sqrt{6 A_{\lambda }+4 \rho _m^2}-3 \rho _m^2\right)+4 \rho _m^4\bigg)^{-1/2}, \\
\tilde{\delta}(\rho_m)\!&=&\!4 \left(6 A_{\lambda }-\frac{3\sqrt{2} \left(2 A_{\lambda }+\rho _m^2\right)}{\sqrt{3 A_{\lambda }+2\rho _m^2}}+2 \rho _m^2\right)\big(3 A_{\lambda } \nonumber\\
                                \!&-&\!\sqrt{6 A_{\lambda }+4 \rho _m^2}+2 \rho _m^2\big)^{-1}.
\end{eqnarray}

\begin{figure*}[htb]
\begin{center}
\subfigure  {\label{ab3}
\includegraphics[width=5.1cm]{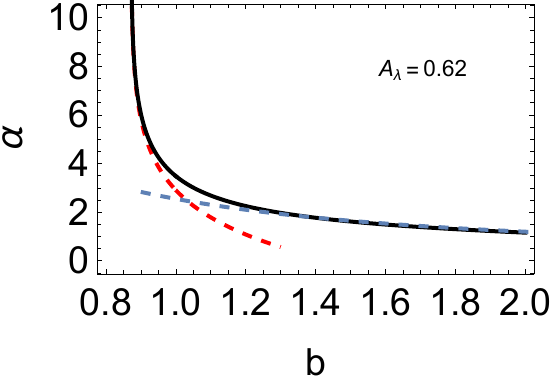}}
\subfigure  {\label{aab}
\includegraphics[width=5cm]{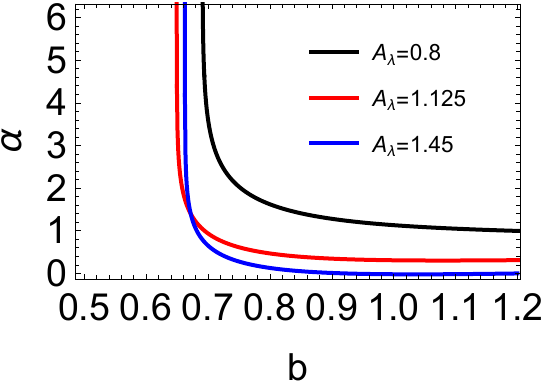}}
\subfigure  {\label{bm}
\includegraphics[width=5.39cm]{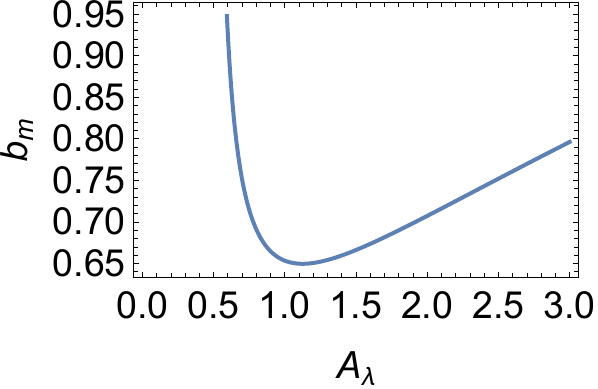}}
\end{center}
\caption{Plots of the exact deflection angle (\ref{exdeflection}) (the black full lines), the corresponding strong deflection limit (\ref{deflection2}) (the red dashed lines) and weak deflection limit (\ref{wedeflection}) (the blue dashed lines) as a function of the impact parameter $b$ for some values of $A_{\lambda}$.}
\label{AA2}
\end{figure*}

Expanding $I_r(\rho_0)$ in powers of $\rho_0-\rho_m$:
\begin{eqnarray}
I_r(\rho_0)=\sum^{\infty}_{j=0}\frac{1}{j!}(\rho_0-\rho_m)^j\int^1_0 \frac{\partial^j \tilde{g}}{\partial \rho^j_0}\Bigg |_{\rho_0=\rho_m}d\tilde{z},
\end{eqnarray}
and focusing on the term of $j=0$, the regular part can be approximated in the strong deflection limit as
\begin{eqnarray}
I_r(\rho_0)\sim I_r(\rho_m)=\int^1_0 \tilde{g}(\tilde{z},\rho_m)d\tilde{z},
\end{eqnarray}
which can be calculated numerically. Figure \ref{AA2} shows the deflection angle of light rays (\ref{deflection2}) in terms of $b$ for the case of $A_{\lambda}>\frac{225}{392}$. Obviously, the strong and weak deflection angles also approximate well to the exact deflection angle, as can be seen in the left panel of Fig.~\ref{AA2}. The middle and right panels of Fig.~\ref{AA2} show that the critical impact parameter has a minimum $b_m=0.6495$ at $A_{\lambda}=1.125$, which indicates that the radius of the photon sphere also reaches a minimum there.

\subsection{Case of $A_{\alpha}=\frac{225}{392}$}

In the case of $A_{\lambda}=\frac{225}{392}$, the unstable photon sphere is located at $x=x_m=0$ or $\rho=\rho_m=\sqrt{A_{\lambda}/2}$, which is marginally unstable since the light rays satisfy $V(x_m)=V'(x_m)=V''(x_m)=V'''(x_m)=0$ and $V''''(x_m)<0$. 

When $A_{\lambda}=\frac{225}{392}$, $\tilde{f}(\tilde{z},\rho_m)$ reduces to
\begin{eqnarray}
\tilde{f}(\tilde{z},\rho_m)&=&\sqrt{\frac{15}{\tilde{z}(2-\tilde{z})}}\Big(4-(\tilde{z}-1)^2 \big(45 (\tilde{z}-1)^2 \nonumber\\
                           &+&28 \sqrt{3 (\tilde{z}-2) \tilde{z}+4} (\tilde{z}-1)+15\big)\Big)^{-1/2},
\end{eqnarray}
and Eqs.~(\ref{c3}), (\ref{c4}) and (\ref{c5}) become
\begin{eqnarray}
c_3(\rho)&=&\frac{855}{196}(\rho_0-\rho_m)^2+\mathcal{O}\big((\rho_0-\rho_m)^3\big), \\
c_4(\rho)&=&\frac{38475}{10976}(\rho_0-\rho_m)+\mathcal{O}\big((\rho_0-\rho_m)^2\big), \\
c_5(\rho)&=&\frac{192375}{307328}+\mathcal{O}\big((\rho_0-\rho_m)\big).
\end{eqnarray}
Then, from Eq.~(\ref{fd}), one can conclude that $\tilde{f}(\tilde{z},\rho_m)$ diverges as $\tilde{z}^{-2/3}$ in this case, and the divergent part of $\tilde{f}(\tilde{z},\rho_m)$ becomes
\begin{eqnarray}
\hat{f}_d(\tilde{z},\rho_m)\equiv\sqrt{\frac{10}{19\tilde{z}^3}},
\end{eqnarray}
which can be integrated out as
\begin{eqnarray}
I_d&=&\int^1_0 2\hat{f}_d(\tilde{z},\rho_0)d\tilde{z}= \left(\frac{4 \sqrt{\frac{10}{19}}}{\sqrt{\tilde{z}}}\right)\Bigg |_{\tilde{z}=0}-4 \sqrt{\frac{10}{19}}, \nonumber\\
     &=& \left(\frac{4 \sqrt{\frac{10}{19}}}{\sqrt{\frac{\rho_0}{\rho_m}-1}}\right)\Bigg |_{\rho_0=\rho_m}-4 \sqrt{\frac{10}{19}}.
\end{eqnarray}

From the expression of the impact parameter in terms of the new radial coordinate, i.e., $b=\sqrt{\frac{\tilde{A}(\rho_0)}{\tilde{C}(\rho_0)}}$, one can obtain
\begin{eqnarray}
b-b_m=\frac{133 \sqrt{\frac{3}{5}}}{16}(\rho_0-\rho_m)^2+\mathcal{O}\big((\rho_0-\rho_m)^3)\big).
\end{eqnarray}
Then, the divergent term $I_d$ as a function of the impact parameter $b$ can be expressed as
\begin{eqnarray}
I_d= \frac{2 \sqrt[4]{\frac{6}{19}} \sqrt{5}}{\sqrt[4]{\frac{b}{b_m}-1}}-4 \sqrt{\frac{10}{19}}.
\end{eqnarray}

\begin{figure*}[htb]
\begin{center}
\subfigure  {\label{ac1}
\includegraphics[width=5.3cm]{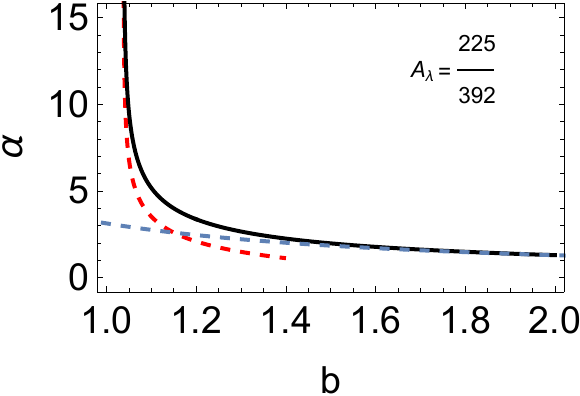}}
\end{center}
\caption{Plots of the exact deflection angle (\ref{exdeflection}) (the black full lines), the corresponding strong deflection limit (\ref{deflection3}) (the red dashed lines) and weak deflection limit (\ref{wedeflection}) (the blue dashed lines) as a function of the impact parameter $b$ for some values of $A_{\lambda}$.}
\label{AA3}
\end{figure*}

Subtracting the divergent part, the regular part is
\begin{eqnarray}
I_r(\rho_m)=\int^1_0 2\big(\tilde{f}(\tilde{z},\rho_m)-\hat{f}_d(\tilde{z},\rho_m)\big)d\tilde{z},
\end{eqnarray}
which can be calculated numerically. Now, the deflection angle of light rays in the strong deflection limit $\rho_0\rightarrow \rho_m$ can be given by
\begin{eqnarray}
\alpha=\frac{\hat{c}}{\sqrt[4]{\frac{b}{b_m}-1}}+\hat{d}+\mathcal{O}[(b-b_m)\ln(b-b_m)],~\label{deflection3}
\end{eqnarray}
where
\begin{eqnarray}
\hat{c}&=&2 \sqrt[4]{\frac{6}{19}} \sqrt{5}\sim 3.35247, \\
\hat{d}&=&-4 \sqrt{\frac{10}{19}}+I_r-\pi\sim -3.24981.
\end{eqnarray}
Figure \ref{AA3} shows that the deflection angles in the strong and weak deflection limits also approximate well to the exact deflection angle.

\section{Observables in the strong deflection limit} ~\label{osdl}

In the previous section, we have obtained the deflection angles as a function of the impact parameter. In this section, we will investigate the observables of this lensing system. We consider a beam of light emanated from a source (S), deflected by a lens (L) of mass $M$, and then arriving at an observer (O). Both the source and the observer are in asymptotically flat spacetime. $D_{\text{LS}}$ stands for the distance between the lens L and the source S, and $D_{\text{OS}}=D_{\text{OL}}+D_{\text{LS}}$ is the distance between the observer O and the source S. Defining the optical axis as the line connecting the lens and the observer, and focusing on a small-angle approximation, i.e., the source, lens, and observer are highly aligned, the lens equation can be written as
\begin{eqnarray}
\beta=\theta-\frac{D_{\text{LS}}}{D_{\text{OS}}}\alpha^{*},~\label{lenseq}
\end{eqnarray}
where $\beta$ and $\theta$ denote the angular position of the source and image with respect to the optical axis, respectively. $\alpha^{*}$ is the effective deflection angle defined by $\alpha^{*}\equiv\alpha-2n\pi$ with $n\in \mathbb{N}$. The deflection angle $\alpha(\theta)$ can be expanded around $\theta=\theta_n^0$ as
\begin{eqnarray}
\alpha(\theta)=\alpha(\theta_n^0)+\frac{d\alpha}{d\theta}\bigg |_{\theta=\theta_n^0}(\theta-\theta_n^0)+\mathcal{O}\big((\theta-\theta_n^0)^2\big),~\label{alex}
\end{eqnarray}
where $\theta_n^0$ is determined by
\begin{eqnarray}
\alpha(\theta_n^0)=2n\pi.~\label{thetan}
\end{eqnarray}

\subsection{$A_{\lambda}\neq \frac{225}{392}$}

In the case of $A_{\lambda}\neq \frac{225}{392}$, both of the deflection angles (\ref{deflection}) and (\ref{deflection2}) are logarithmic divergence in the strong deflection limit, and they can be rewritten as a function of $\theta$:
\begin{eqnarray}
\alpha(\theta)&=&-\tilde{\bar{a}}\ln\left(\frac{\theta}{\theta_{\infty}}-1\right)+\tilde{\bar{u}} \nonumber\\
                     &+&\mathcal{O}\left(\left(\frac{\theta}{\theta_{\infty}}-1\right)\ln\left(\frac{\theta}{\theta_{\infty}}-1\right)\right),~\label{defthe}
\end{eqnarray}
where $\tilde{\bar{X}}$ denotes $\bar{X}$ or $\tilde{X}$, and we have used the first-order approximations $\theta\approx b/ D_{\text{OL}}$ and $\theta_{\infty}\approx b_m/ D_{\text{OL}}$. From the above expression, one can easily obtain
\begin{eqnarray}
\frac{d\alpha}{d\theta}\bigg |_{\theta=\theta_n^0}=-\frac{\tilde{\bar{a}}}{\theta_n^0-\theta_{\infty}}.~\label{aldth}
\end{eqnarray}
Solving Eq.~(\ref{defthe}) with $\alpha(\theta_n^0)=2n\pi$, we find
\begin{eqnarray}
\theta_n^0=\left(1+\text{e}^{\frac{\tilde{\bar{u}}-2n\pi}{\tilde{\bar{a}}}}\right).~\label{theta0}
\end{eqnarray}
Then, from Eqs.~(\ref{alex}), (\ref{aldth}) and (\ref{theta0}), the effective deflection angle $\alpha^{*}$ can be calculated as
\begin{eqnarray}
\alpha^{*}(\theta_n)=\frac{\tilde{\bar{a}}}{\theta_{\infty}\text{e}^{\frac{\tilde{\bar{u}}-2n\pi}{\tilde{\bar{a}}}}}(\theta_n^0-\theta_n).~\label{effdefan}
\end{eqnarray}

After inserting the effective deflection angle (\ref{effdefan}) into the lens equation (\ref{lenseq}), the angular position of the $n$-th image can be calculated as
\begin{eqnarray}
\theta_n(\beta)=\theta_n^0+\frac{\theta_{\infty}\text{e}^{\frac{\tilde{\bar{u}}-2n\pi}{\tilde{\bar{a}}}}D_{\text{OS}}(\beta-\theta_n^0)}{\tilde{\bar{a}}D_{\text{LS}}},
\end{eqnarray}
where the second term is a small correction to $\theta_n^0$. In the case of the observer, the lens, and the source perfectly aligned in a line, i.e., $\beta=0$, the point images become an infinite series of concentric rings with angular position
\begin{eqnarray}
\theta_{\text{E}n}\equiv \theta_n(0)=\left(1-\frac{\theta_{\infty}\text{e}^{\frac{\tilde{\bar{u}}-2n\pi}{\tilde{\bar{a}}}}D_{\text{OS}}}{\tilde{\bar{a}}D_{\text{LS}}}\right)\theta_n^0,
\end{eqnarray}
which are the so-called Einstein rings.

With the approximation of $\theta_n\sim\theta_n^0$, the magnification of the $n$-th relativistic image can be given by
\begin{eqnarray}
\mu_n\equiv\frac{1}{\frac{\beta}{\theta}\frac{d\beta}{d\theta}}\Bigg |_{\theta=\theta_n^0}\simeq\frac{\theta_{\infty}^2D_{\text{OS}}}{\tilde{\bar{a}}\beta D_{\text{LS}}}\text{e}^{\frac{\tilde{\bar{u}}-2n\pi}{\tilde{\bar{a}}}}\left(1+\text{e}^{\frac{\tilde{\bar{u}}-2n\pi}{\tilde{\bar{a}}}}\right).
\end{eqnarray}
Obviously, the magnification gets exponentially suppressed with increasing $n$, which means that the first image is the brightest one and the brightness of other relativistic images are highly demagnified unless the lens is highly aligned with the source, i.e., $\beta\sim 0$.

In the simplest situation, we consider that only the outermost relativistic image 􏰆$\theta_1$ can be resolved separately and all other relativistic images are packed together at the limiting angular position $\theta_{\infty}=b_m/D_{\text{OL}}$. Then, the angular separation $s$ between the outermost one and the others can be defined by
\begin{eqnarray}
s\equiv \theta_1-\theta_{\infty}\sim \theta_1^0-\theta_{\infty}^0=\theta_{\infty}\text{e}^{\frac{\tilde{\bar{u}}-2\pi}{\tilde{\bar{a}}}}, ~\label{ss}
\end{eqnarray}
and the quotient of the flux of the outermost relativistic image to that of all other relativistic images is
\begin{eqnarray}
r\equiv \frac{\mu_1}{\sum^{\infty}_{n=2}\mu_n} \sim \frac{\big(\text{e}^{\frac{4\pi}{\tilde{\bar{a}}}}-1\big)\big(\text{e}^{\frac{2\pi}{\tilde{\bar{a}}}}+\text{e}^{\frac{\tilde{\bar{u}}}{\tilde{\bar{a}}}}\big)}{\text{e}^{\frac{4\pi}{\tilde{\bar{a}}}}+\text{e}^{\frac{2\pi}{\tilde{\bar{a}}}}+\text{e}^{\frac{\tilde{\bar{u}}}{\tilde{\bar{a}}}}}, ~\label{rr}
\end{eqnarray}
where the cumulative flux from all other relativistic images except the outermost one is
\begin{eqnarray}
\sum^{\infty}_{n=2}\mu_n \sim \frac{\theta_{\infty}^2D_{\text{OS}}\big(\text{e}^{\frac{4\pi}{\tilde{\bar{a}}}}+\text{e}^{\frac{2\pi}{\tilde{\bar{a}}}}+\text{e}^{\frac{\tilde{\bar{u}}}{\tilde{\bar{a}}}}\big)\text{e}^{\frac{\tilde{\bar{u}}-4\pi}{\tilde{\bar{a}}}}}{\beta\tilde{\bar{a}}D_{\text{LS}}\big(\text{e}^{\frac{4\pi}{\tilde{\bar{a}}}}-1\big)}.
\end{eqnarray}

\subsection{$A_{\lambda}=\frac{225}{392}$}

In the case of $A_{\lambda}=\frac{225}{392}$, the deflection angle as a function of $\theta$ can be derived as
\begin{eqnarray}
\alpha(\theta)=\frac{\hat{c}}{\left(\frac{\theta}{\theta_{\infty}}-1\right)^{1/4}}+\hat{d},~\label{alphthe}
\end{eqnarray}
which yields
\begin{eqnarray}
\frac{d\alpha}{d\theta} \bigg |_{\theta=\theta_n^0}=-\frac{\hat{c}}{4\theta_{\infty}}\left(\frac{\theta_n^0}{\theta_{\infty}}-1\right)^{-5/4}.
\end{eqnarray}
Then, from Eqs.~(\ref{thetan}) and (\ref{alphthe}), one can obtain
\begin{eqnarray}
\theta_n^0=\left[1+\left(\frac{\hat{c}}{2n\pi-\hat{d}}\right)^4\right]\theta_{\infty},
\end{eqnarray}
and the effective deflection angle can be calculated as
\begin{eqnarray}
\alpha^{*}(\theta_n)=\frac{(2n\pi-\hat{d})^5}{4\theta_{\infty}\hat{c}^4}(\theta_n^0-\theta).
\end{eqnarray}

Inserting the effective deflection angle into the lens equation (\ref{lenseq}), the angular positions of the relativistic images are solved as
\begin{eqnarray}
\theta_n(\beta)= \theta_n^0+\frac{4\hat{c}^4D_{\text{OS}}\theta_{\infty}(\beta-\theta_n^0)}{D_{\text{LS}}(2n\pi-\hat{d})^5}.~\label{thetan2}
\end{eqnarray}
Then, the relativistic Einstein rings are given by
\begin{eqnarray}
\theta_{\text{E}n}(0)=\left[1-\frac{4\hat{c}^4D_{\text{OS}}\theta_{\infty}}{D_{\text{LS}}(2n\pi-\hat{d})^5}\right]\theta_n^0.
\end{eqnarray}

With the approximation of $\theta_n\sim\theta_n^0$, the magnification of the $n$-th image is
\begin{eqnarray}
\mu_n\simeq \frac{4\theta_{\infty}^2\hat{c}^4D_{\text{OS}}F_n}{\beta D_{\text{LS}}},
\end{eqnarray}
where
\begin{eqnarray}
F_n=\frac{1+\left(\frac{\hat{c}}{2n\pi-\hat{d}}\right)^4}{(2n\pi-\hat{d})^5}.
\end{eqnarray}

The separation of the angular positions between the outermost relativistic image and the others is
\begin{eqnarray}
s=\theta_1-\theta_{\infty}=\left(\frac{\hat{c}}{2\pi-\hat{d}}\right)^4\theta_{\infty},~\label{ssc}
\end{eqnarray}
and the luminosity ratio of the flux of the outermost one to that of all others is
\begin{eqnarray}
r=\frac{\mu_1}{\sum^{\infty}_{n=2}\mu_n}\sim \frac{F_1}{\sum^{\infty}_{n=2}F_n}=9.95368,~\label{rrc}
\end{eqnarray}
where
\begin{eqnarray}
F_1\sim 1.28957\times 10^{-5},
\end{eqnarray}
and
\begin{eqnarray}
\sum^{\infty}_{n=2}F_n\sim 1.29557\times 10^{-6}.
\end{eqnarray}

From the observation of the relativistic images, one can obtain the limiting angular position $\theta_{\infty}$, the angular separation $s$, and the flux ratio $r$. By inverting Eqs.~(\ref{ss}) and (\ref{rr}) or  Eqs.~(\ref{ssc}) and (\ref{rrc}), one then obtains the coefficients of the strong deflection angle, which contain the information about the parameters of the lensing black hole.

\subsection{Applications: Black holes in our Galaxy and M87}

\begin{figure*}[htb]
\begin{center}
\subfigure  {\label{theta}
\includegraphics[width=5.2cm]{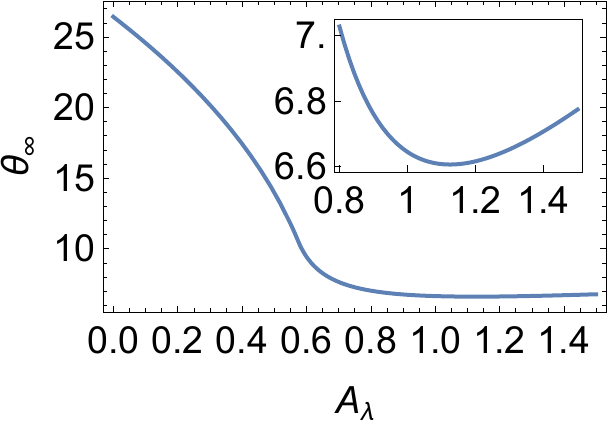}}
\subfigure  {\label{s}
\includegraphics[width=5.3cm]{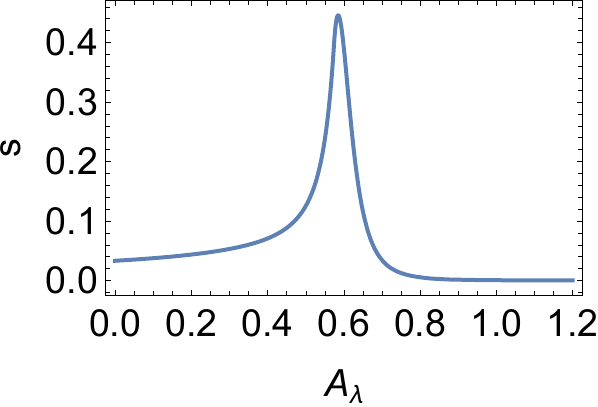}}
\subfigure  {\label{r}
\includegraphics[width=5.29cm]{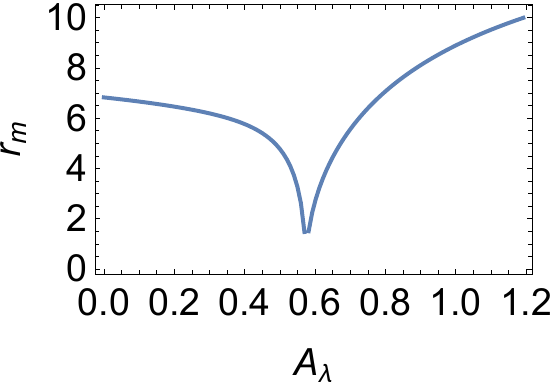}}
\end{center}
\caption{Plots of the observables $\theta_{\infty}$ (left), $s$ (middle), and $r$ in magnitudes ($r_m=2.5\log r$, right) as a function of $A_{\lambda}$ for the supermassive black hole at the center of our Galaxy.}
\label{obfig}
\end{figure*}

\begin{figure*}[htb]
\begin{center}
\subfigure  {\label{theta}
\includegraphics[width=5.2cm]{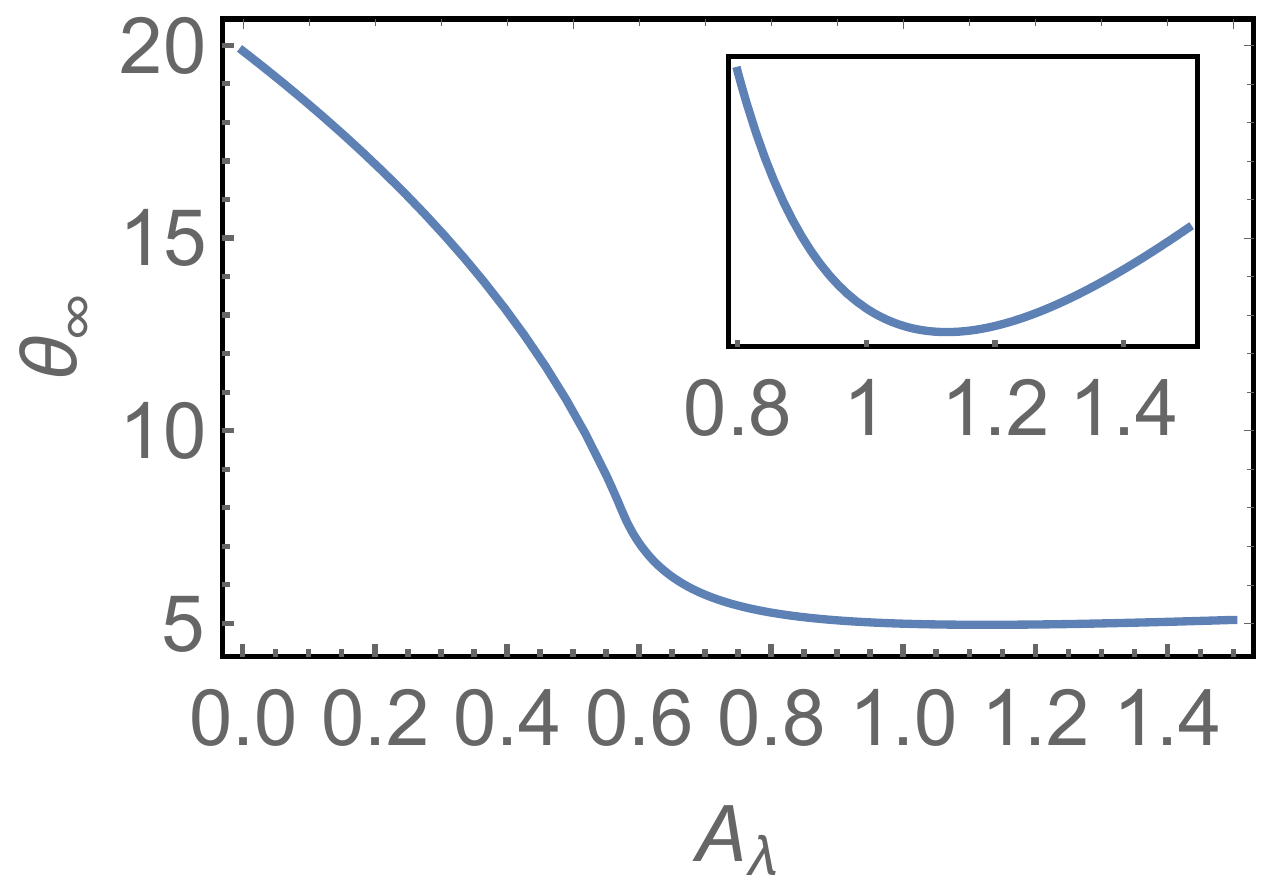}}
\subfigure  {\label{s}
\includegraphics[width=5.3cm]{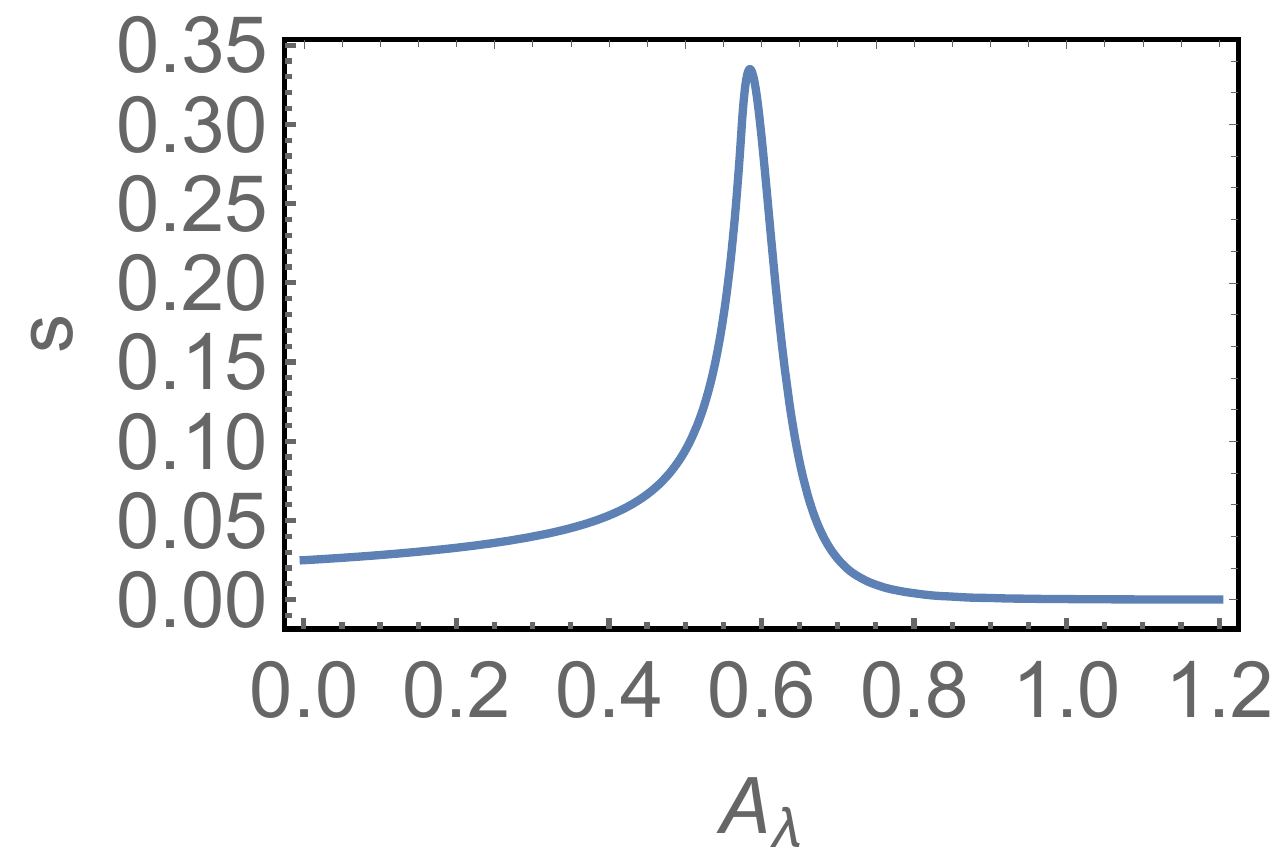}}
\subfigure  {\label{r}
\includegraphics[width=5.29cm]{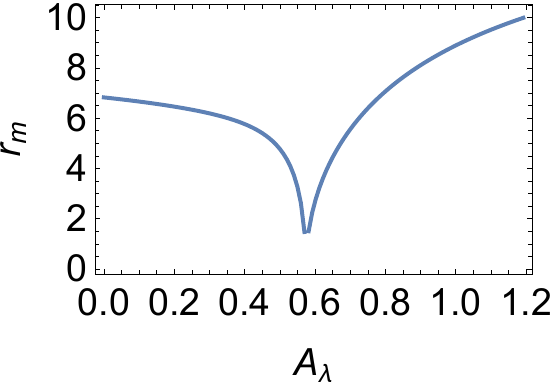}}
\end{center}
\caption{Plots of the observables $\theta_{\infty}$ (left), $s$ (middle), and $r$ in magnitudes ($r_m=2.5\log r$, right) as a function of $A_{\lambda}$ for the supermassive black hole at the center of M87.}
\label{obfig2}
\end{figure*}

To evaluate the observables defined above,  {we respectively take the black holes with mass $M=4.3\times 10^6 M_{\odot}$ and $6.5\times 10^9 M_{\odot}$ at the centers of our Galaxy \cite{DoL27} and $M87$ \cite{EHTI} as examples} and apply the strong deflection limit. The distances between the observer and the lens are taken as $D_{OL} = 8.35$~kpc and {$16.8$ Mpc, respectively}. The numerical estimates for the observables with different $A_{\lambda}$ are given in Tables \ref{obda} and \ref{obda2}. 
It can be seen that the limiting angular position $\theta_{\infty}$ decreases rapidly with increasing $A_{\lambda}$ until $A_{\lambda}\sim 0.7$ for both black holes, indicating that the limiting relativistic images are closer to the optical axis than the Schwarzschild case, and then there is a little bounce at $A_{\lambda}=1.125$, as shown in the left panels of Figs.~\ref{obfig} and \ref{obfig2}. {Since the independences of the black hole mass $M$ and the distance $D_{OL}$, the relative magnifications are the same for both black holes as shown in Tables \ref{obda} and \ref{obda2} or the right panels of Figs.~\ref{obfig} and \ref{obfig2}.} Obviously, the relative magnifications in magnitudes $r_m$ first decrease and then increase as $A_{\lambda}$ increases, which suggests that the outermost relativistic image first becomes fainter and then brighter compared to the others. Besides, we also find that there is a maximum of the observable $s$ at $A_{\lambda}=0.58$ for both black holes, as shown in the middle panels of Figs.~\ref{obfig} and \ref{obfig2}, which means that the angular separation between the outermost relativistic image and the others first becomes wider and then decreases with increasing $A_{\lambda}$. This feature can be easily recognized in the astronomical observation. What's more, since the lensing parameters are associated with a wormhole when $A_{\lambda}\geqslant\frac{1}{2}$, the maximum of the angular separation $s$ signals that a wormhole has formed. {Recently, from the image of the black hole shadow \cite{Brahma181301}, the perihelion advance, and Shapiro time Delay \cite{Liu10202}, the tightest constraint on $A_{\lambda}$ is $0<A_{\lambda}<4\times 10^{-6}$. With the maximum $A_{\lambda}=4\times 10^{-6}$, the observables for the black hole Sgr$\text{A}^{\ast}$ are $\theta_{\infty}=26.4232322$, $s=0.0331672$ and $r_m=6.8218757$, which correspond to a $4.46\times10^{-6}$ deviation from the case of the Schwarzschild black hole at best.}

\begin{table*}
\caption{Numerical estimations for the main observables for the black hole at the center of our Galaxy.}
\begin{tabular}{|l|c|c|c|c|c|c|c|c|c|c|}
  \hline 
  Observable  & Schwarzschild & \multicolumn{8}{|c|}{Loop Quantum Black Hole} \\
    \cline{3-10}
    & ($A_{\lambda}=0$)  & $A_{\lambda}=0.1$ & $A_{\lambda}=0.3$ & $A_{\lambda}=0.5$ & $A_{\lambda}=\frac{225}{392}$ & $A_{\lambda}=0.7$ & $A_{\lambda}=0.9$ & $A_{\lambda}=1.1$ & $A_{\lambda}=1.2$\\
    \hline
  $\theta_\infty$ ($\mu$arcsec) & 26.42 & 24.57 & 20.18 & 13.98 & 10.55 & 7.64 & 6.75 & 6.60 & 6.61 \\
  $s$ ($\mu$arcsec) & 0.0331 & 0.0375 & 0.0529 & 0.1252 & 0.3961 & 0.0343 & 0.0014 & 0.0002 &  0.0001 \\
  $r_m$ (magnitude) & 6.82 & 6.65 & 6.18 & 4.77 & 2.49 & 5.54 & 8.10 & 9.51 & 10.02 \\
\hline
\end{tabular}
\label{obda}
\end{table*}

\begin{table*}
\caption{Numerical estimations for the main observables for the black hole at the center of M87.}
\begin{tabular}{|l|c|c|c|c|c|c|c|c|c|c|}
  \hline 
  Observable  & Schwarzschild & \multicolumn{8}{|c|}{Loop Quantum Black Hole} \\
    \cline{3-10}
    & ($A_{\lambda}=0$)  & $A_{\lambda}=0.1$ & $A_{\lambda}=0.3$ & $A_{\lambda}=0.5$ & $A_{\lambda}=\frac{225}{392}$ & $A_{\lambda}=0.7$ & $A_{\lambda}=0.9$ & $A_{\lambda}=1.1$ & $A_{\lambda}=1.2$\\
    \hline
  $\theta_\infty$ ($\mu$arcsec) & 19.85 & 18.46 & 15.16 & 10.50 & 7.92 & 5.74 & 5.07 & 4.96 & 4.97 \\
  $s$ ($\mu$arcsec) & 0.0249 & 0.0281 & 0.0398 & 0.0941 & 0.2975 & 0.0258 & 0.0011 & 0.0001 &  0.00009 \\
  $r_m$ (magnitude) & 6.82 & 6.65 & 6.18 & 4.77 & 2.49 & 5.54 & 8.10 & 9.51 & 10.02 \\
\hline
\end{tabular}
\label{obda2}
\end{table*}

\section{Conclusion} ~\label{con}

In this paper, we investigated the deflection angles of light rays by a black hole in effective loop quantum gravity (LQG) in the strong and weak deflection limits. It was found that the weak deflection angle is consistent with the Schwarzschild black hole  at the lowest order, and the corrections originating from quantum effects make a negative contribution to the weak deflection angle. The primary quantum correction is of the same order as the spin or charge of a black hole.

Then, we investigated the bending angle in the strong deflection limit. It was found that the LQG black hole reduces to a Schwarzschild black hole for vanishing $A_{\lambda}$, a regular black hole for $0<A_{\lambda}<\frac{1}{2}$, a one-way traversable wormhole with a null throat for $A_{\lambda}=\frac{1}{2}$, and a traversable wormhole with two-way throat for $A_{\lambda}>\frac{1}{2}$. Besides, if $M=0$, the LQG black hole disappears and the spacetime is flat as expected, while  the spacetime of the black-bounce black hole reduces to an Ellis wormhole with vanishing $M$, which seems unnatural \cite{Tsukamoto024033,Simpson042}. There are two unstable photon spheres for $0<A_{\lambda}<\frac{225}{392}$ located at $r=\pm r_m$, respectively. When $A_{\lambda}=\frac{225}{392}$, these two unstable photon spheres merge into one marginally unstable photon sphere. It was found that the behavior of the strong deflection angle exhibits logarithmic divergence in terms of the impact parameter for $A_{\lambda}\neq \frac{225}{392}$ and nonlogarithmic divergence for $A_{\lambda}= \frac{225}{392}$.

Next, we calculated the observables of this lensing system, such as the angular positions of the relativistic images, the angular separation between the outermost relativistic image and the others, and also their flux ratio. These observables were estimated numerically, using the black holes {at the centers of our Galaxy and M87 as examples}. We showed that {the observables for both black holes have the similar behaviors. To be specific,} the limiting angular position $\theta_{\infty}$ and the relative magnification in magnitudes $r_m$ decrease with $A_{\lambda}$ for $0\leqslant A_{\lambda}\leqslant\frac{225}{392}$. However, when $A_{\lambda}>\frac{225}{392}$, there is a little bounce at $A_{\lambda}=1.125$ in the limiting angular position, and the relative magnification in magnitudes $r_m$ increases with $A_{\lambda}$, suggesting that the outermost relativistic image gradually becomes brighter relative to the others. Besides, it is interesting to find that there is a maximum in the angular separation at $A_{\lambda}=0.58$, which means that the angular separation first becomes wider and then decreases with increasing $A_{\lambda}$. The maximum is reached after the wormhole is formed, which can be considered as a signal for the formation of a wormhole. It is expected that these features will be used to distinguish LQG from general relativity.

\acknowledgments{
This work was supported by scientific research programs funded by the National Natural Science Foundation of China (Grants No. 11975072, 11835009, 11875102, and 11690021), the Liaoning Revitalization Talents Program (Grant No. XLYC1905011), the Fundamental Research Funds for the Central Universities (Grant No. N2005030), the National Program for Support of Top-Notch Young Professionals (Grant No. W02070050), the National 111 Project of China (Grant No. B16009), and Research Foundation of Education Bureau of Shaanxi Province, China (Grant No. 19JS009).
}

\end{document}